\documentclass[sigconf,screen]{acmart}
\acmYear{2025}\copyrightyear{2025}
\setcopyright{rightsretained}
\acmConference[Preprint version, PROMISE '25]{21st International Conference on Predictive Models and Data Analytics in Software Engineering}{June 26, 2025}{Trondheim, Norway}
\acmBooktitle{21st International Conference on Predictive Models and Data Analytics in Software Engineering (PROMISE '25), June 26, 2025, Trondheim, Norway}
\acmDOI{10.1145/3727582.3728686}
\acmISBN{979-8-4007-1594-5/25/06}

\setcopyright{none}

\usepackage{algorithm}
\usepackage{algorithmic}
\usepackage{graphicx}
\usepackage{xcolor}
\usepackage{diagbox}
\usepackage{tcolorbox}
\usepackage{csquotes}
\usepackage{framed}
\usepackage{multirow}

\tcbuselibrary{skins}
\tcbset{RQsboxstyle/.style={
    width=\columnwidth,       
    colframe=gray,            
    colback=gray!10,          
    coltitle=black,           
    boxrule=0.1mm,            
    arc=1mm,                  
    auto outer arc,           
    boxsep=1mm,               
    drop shadow={black!50!white},  
    leftrule=1.5mm
}}

\tcbset{yellowboxstyle/.style={
    width=\columnwidth,       
    colframe=yellow!40!black,          
    colback=yellow!10,        
    coltitle=black,           
    boxrule=0.1mm,            
    arc=1mm,                  
    auto outer arc,           
    boxsep=1mm,               
    drop shadow={black!50!white} 
}}

\begin{document}

\title{
Security Bug Report Prediction Within and Across Projects: A Comparative Study of BERT and Random Forest
}

\author{Farnaz Soltaniani}
\email{farnaz.soltaniani@tu-clausthal.de}
\affiliation{%
  \institution{Technische Universität Clausthal}
  \country{Germany}
}
\author{Mohammad Ghafari}
\email{mohammad.ghafari@tu-clausthal.de}
\affiliation{%
  \institution{Technische Universität Clausthal}
  \country{Germany}
}
\author{Mohammed Sayagh}
\email{mohammed.sayagh@etsmtl.ca}
\affiliation{%
  \institution{ETS - Quebec University}
  \country{Canada}
}

\begin{abstract}

Early detection of security bug reports (SBRs) is crucial for preventing vulnerabilities and ensuring system reliability.
While machine learning models have been developed for SBR prediction, their predictive performance still has room for improvement.
In this study, we conduct a comprehensive comparison between BERT and Random Forest (RF), a competitive baseline for predicting SBRs. 
The results show that 
RF outperforms BERT with a 34\% higher average G-measure for within-project predictions.
Adding only SBRs from various projects improves both models' average performance.
However, including both security and nonsecurity bug reports significantly reduces RF's average performance to 46\%, while boosts BERT to its best average performance of 66\%, surpassing RF. 
In cross-project SBR prediction, BERT achieves a remarkable 62\% G-measure, which is substantially higher than RF.

\end{abstract}



\maketitle

\section{Introduction}

Software systems are inherently exposed to security risks—whether built with emerging technologies like WebAssembly~\cite{Quentin22}, established platforms like Android~\cite{Ghafari2017}, or deployed in mission-critical environments~\cite{Wetzels2023}.
AI-assisted development tools have accelerated code production, but they are also susceptible to generating vulnerable code~\cite{Bruni2025}, further amplifying the need for vigilant and consistent attention to security risks.

Bug reports (BRs) are the primary means to inform development teams of a problem or defect found in software.
Nevertheless, studies have shown that security bug reports (SBRs) often progress slowly, with many remaining unresolved for extended periods~\cite{Noah2022}.
Prioritizing SBRs over Non-Security bug reports (NSBRs) is critical to addressing vulnerabilities promptly and preventing potential exploitation. Yet, limited security knowledge among mainstream developers often leads to SBRs being misclassified as NSBRs during reporting~\cite{5463340}, delaying vulnerability detection and mitigation.

Researchers have built machine learning techniques to classify SBRs from NSBRs, primarily based on the description field of bug reports~\cite{FARSEC, 9371393, Ohiraetal, DeeplearningBasedSoftwareBugClassification2024, 10.1145/3643991.3644903}. 

However, they often perform poorly due to class imbalance and limited semantic understanding, especially in large, diverse datasets. 
Despite efforts to enhance SBR prediction, distinguishing minority classes remains a challenge \cite{reducefalsepositives}.

BERT \cite{BERTTransformers}, 
a pre-trained transformer-based model, offers exceptional semantic understanding and has achieved state-of-the-art performance in software issue classification~\cite{10.1145/3528588.3528659, 10.1109/TSE.2022.3178469, izadi2021predictingobjectivepriorityissue, vaswani2023attentionneed}.
However, there is not much evidence on BERT performance for SBR prediction. Therefore, in this paper, we extensively compare the performance of BERT in predicting SBRs to Random Forest (RF), a strong baseline \cite{9371393}. We assess their performance for within and across projects SBRs prediction, assess the models in scenarios where project-specific data is unavailable, and examine how additional bug reports impact their prediction capabilities.
In particular, we answer the following research questions:

\textbf{RQ1}: How effectively can BERT predict SBRs compared to the state-of-the-art RF?

Based on the data sets used by Wu et al. \cite{9371393}, we evaluate the effectiveness of BERT and RF in predicting SBRs. 
Using a within-project prediction (WPP) approach, in which models are trained and tested on data from the same software project, we train models on historical bug reports to predict future, unseen ones.
Our findings show that RF outperforms BERT, with a 34\% higher average G-measure.
However, in the largest dataset where both models demonstrate their best performance, BERT achieves a G-measure of 83\%, which is higher than RF's 75\%.

\textbf{RQ2}: What is the impact of augmenting data on the performance of security bug report predictors?

We evaluate the performance of BERT and RF in predicting SBRs by incorporating additional data from diverse real-world projects. We train our model using historical BRs from one selected project combined with data from other projects.
We initially incorporate SBRs and later extend the training data to include NSBRs as well.
We evaluate the performance of models on the most recent BRs from the chosen project to determine the impact of data increase. 

Incorporating SBRs from other projects enhances the average G-measures of both models, i.e., 70\% and 59\% for RF and BERT, respectively.
We note that adding SBRs improves BERT more than RF.
Nonetheless, when we add both SBRs and NSBRs from other projects, the average G-measure for RF drops significantly to 46\%, whereas it increases to 66\% for BERT, unleashing its best performance.
In comparison with the initial WPP (ie., RQ1), the addition of BRs results in 43.48\% improvement for BERT.

\textbf{RQ3}: How do BERT and RF models compare for cross-project prediction?

We assess the ability of BERT and RF to predict SBRs in a target project with limited or no BRs using knowledge from other projects, known as cross-project prediction (CPP).
We start with a single project and progressively include BRs from additional projects, evaluating each model's ability to predict SBRs on the target project. Results show that building a model based on one project may not effectively predict SBRs in another project. However, using multiple projects for training, BERT achieves an average G-measure of 62\%, which is substantially higher than that of RF's 38\%.

In summary, this study presents the first comprehensive comparison of BERT versus the state-of-the-art RF model for SBR prediction. 
For WPP, RF outperforms BERT with an average G-measure that is 34\% higher. Nevertheless, we find that adding more data enhances BERT's performance, whereas RF performs better with only SBRs and experiences a performance drop when all BRs are included. Notably, BERT achieves an average G-measure of 62\% for cross-project SBR prediction, which is significantly superior to that of RF. To support future research, we have publicly shared our replication package.\footnote{\url{https://doi.org/10.5281/zenodo.15240582}}

The paper is structured as follows.
Section \ref{sec:RelatedWork} presents related work. Section \ref{sec:Methodology} describes the methodology used. Section \ref{sec:ExperimentalSetup} explains the experimental setup, and Section \ref{sec:EXPERIMENTRESULTS} details the results for each research question. We discuss the threats to the validity of our study in Section \ref{sec:ThreatsToValidity}. Finally, Section \ref{sec:Conclusion} concludes our findings.

\section{Related Work}
\label{sec:RelatedWork}

Researchers have proposed text-based machine learning approaches to improve the classification of bug reports as either SBRs or NSBRs in bug-tracking systems \cite{Zaher2021, 6473768}. However, the scarcity of SBRs leads to a significant class imbalance, resulting in sparse feature representations that degrade the performance of traditional machine learning models.
To address this issue, researchers suggest preprocessing datasets before applying machine learning algorithms \cite{FARSEC, shu2019better, LTRWES}. This preprocessing may involve filtering out NSBRs that contain security-related keywords \cite{FARSEC}, using content-based filtering to minimize the risk of misclassifying NSBRs as SBRs \cite{LTRWES}, utilizing the Synthetic Minority Oversampling Technique (SMOTE) and its variant, SMOTUNED \cite{shu2019better}, or applying a k-means clustering approach. The k-means method maintains diversity among NSBRs by grouping data based on structure and selecting the samples nearest to each cluster's centroid \cite{CASMS}.

To exemplify, Peters et al. \cite{FARSEC} introduced FARSEC, a framework designed to improve SBR prediction by filtering out misleading NSBRs. The framework identifies the top 100 terms relevant to SBRs and scores NSBRs based on the frequency and significance of these terms. NSBRs with scores above a specific threshold (e.g., 0.75) are filtered out as likely misclassifications. To implement this, seven distinct filters were developed and applied to the training set, creating seven new training sets for independent model fitting. Subsequently, five machine learning algorithms, including RF, were used to predict the probability of a bug report being security-related. 

Wu et al. \cite{9371393} built upon the work of Peters et al. \cite{FARSEC} to improve label accuracy across the same five datasets. They employed various classification methods, including the FARSEC approach with the \textit{farsectwo} filter, which involves applying the Graham version by multiplying the frequency by two, achieving 80\% of the best results reported by Peters et al.
Their experiments demonstrated that using the same prediction models as Peters et al. \cite{FARSEC}, simple text classification methods outperformed both the hyperparameter-tuned models and the data preprocessing techniques applied by Shu et al. \cite{shu2019better}. Notably, RF exhibited the best overall performance, both with and without the use of FARSEC, highlighting its effectiveness for this task. However, previous methods had some limitations.

Researchers have also investigated cross-project learning for SBR prediction.
Gegick et al. \cite{5463340} highlighted that models trained on one dataset may not generalize well to systems with differing SBRs, whereas Scandariato et al. \cite{6860243} suggested that certain single-application models can effectively predict vulnerabilities across different applications.
Peters et al. \cite{FARSEC} found that CPP models often outperform WPP models, especially when SBRs are scarce, demonstrating their value for projects with limited labeled data.

\begin{table*}[htb] 
\footnotesize
    \centering
\setlength{\tabcolsep}{16.5pt} 
\renewcommand{\arraystretch}{1.3}
    \caption{Dataset-related information, where SBRs (\%) indicates the proportion of security bug reports}
    \scalebox{1.0}{
    \begin{tabular}{l l c c c c}
        \hline 
        Project & Description & BRs & SBRs & NSBRs & SBRs (\%) \\
        \hline \hline
        Chromium & Web browser called Chrome & 41,940 & 808 & 41,132 & 1.92 \\
        
        Derby & A relational database management system & 1,000 & 179 & 821 &  17.9 \\
        
        Camel & A rule-based routing and mediation engine & 1,000 & 74 & 926 & 7.4 \\
        
        Ambari & Hadoop management web UI backed by its RESTful APIs & 1,000 & 56 & 944 & 5.6 \\
        
        Wicket & Component-based web application framework for Java programming & 1,000 & 47 & 953 & 4.7 \\
        \hline
    \end{tabular}}
    \label{tab:datasetsDetails}
\end{table*}

To sum up, unlike previous studies, our work primarily focuses on BERT, which exhibits exceptional semantic understanding in text classification. 
We provide a comprehensive comparison between BERT and RF, a strong baseline for SBR prediction. 
Our evaluation spans both WPP and CPP scenarios and investigates the impact of augmenting training data from multiple sources on model performance.

\section{Methodology}
\label{sec:Methodology}

SBR prediction automates manual classification, reducing time, effort, and cost.
Recent advancements in hardware, particularly the accessibility of GPUs, have enabled transformer-based Large Language Models (LLMs) \cite{vaswani2023attentionneed} to drive significant progress in Natural Language Processing (NLP). These models leverage self-attention mechanisms to process input sequences more effectively, capturing long-range dependencies between tokens. A key factor in their success is the pre-training phase, where they learn language structures from vast amounts of unlabeled data, leading to state-of-the-art performance across various NLP tasks. Based on the transformer architecture defined by Pan et al. \cite{Pan_2024}, BERT is an encoder-only model, particularly well-suited for sentence-level tasks.

In this study, we first compare the performance of BERT and RF in SBRs prediction using the BRs of the same software project.
We then expand our research by evaluating their predictive performance by augmenting additional data from other projects with a twofold augmentation strategy: (1) increasing SBR instances and (2) adding BRs to maintain diversity. Additionally, we evaluate model generalizability by training on specific projects and testing on an external project.

\subsection{Model selection}
Among encoder-only models, BERT \cite{BERTTransformers} is trained using two key objectives: masked language modeling (MLM) and next-sentence prediction (NSP). In MLM, certain tokens in the input sequence are masked and predicted based on the surrounding context, while NSP determines whether two sentences appear consecutively in the original text. Fine-tuning BERT involves adding a simple classification layer to the pre-trained model, allowing all parameters to be adjusted for specific tasks. This transfer learning approach makes BERT highly adaptable to domain-specific datasets and has demonstrated state-of-the-art performance in software issue classification, effectively distinguishing bugs from improvements, new features, and questions \cite{10.1145/3528588.3528659, 10.1109/TSE.2022.3178469}.

We use the \textit{BertForSequenceClassification} model from Hugging Face \cite{wolfetal2020transformers}, a standard BERT model with an additional layer for document classification. To prepare our dataset for the pre-trained BERT, we tokenize our bug reports' descriptions using the \textit{BertTokenizer} from Hugging Face, specifically the ``\textit{bert-base-uncased}'' case-insensitive version, treating words like ``issue'' and ``Issue'' as identical. The tokenized text is converted into smaller tokens mapped to vocabulary indices. Padding is dynamically applied to sequences within each batch using \textit{DataCollatorWithPadding}, and attention masks are generated to distinguish real tokens from padding, allowing the model to focus only on meaningful inputs. Finally, the token IDs and attention masks are used to create a TensorDataset, which is fed into a DataLoader configured with random sampling and a batch size of 32 for training and evaluation \cite{devlinetal2019bert}.

For comparison, we adopt RF, which Wu et al. \cite{9371393} identified as the best baseline method for SBR prediction. To generalize their approach, we also follow their methodology for building prediction models using the FARSEC datasets and evaluate how FARSEC performs with BERT.
We replicated the FARSEC approach using the replication package from Peters et al. However, due to some missing parameters, we had to implement fixes in the code to make it functional. We used the same parameters as outlined in the detailed study by Wu et al., including a count of security-related keywords set at 100, the parameter tuning ranges, and the \textit{farsectwo} filtering method. Throughout our study, we refer to this method simply as FARSEC, in line with the terminology used in the literature. Wu et al. utilized the \textit{sklearn} package for implementing RF, which has since been deprecated. In our implementation, we used the \textit{scikit-learn} library, specifically its \textit{RandomForestClassifier}, for model training and predictions. To enhance model performance, we employed a differential evolution strategy for hyperparameter tuning, allowing for optimized selection of hyperparameters such as the number of trees, maximum tree depth, minimum samples per split, minimum samples per leaf, and the maximum number of features considered at each split.

\subsection{Data Augmentation}
We explore two distinct evaluations for predicting future unseen SBRs: WPP and CPP. 

\subsubsection{Within-Project Prediction (WPP)}
Prediction models can be trained and tested using data from the same project, known as WPP.  
In WPP, training and testing prediction models are applied in a single system using cross-validation techniques \cite{shu2019better}; or in several releases of a system, building prediction models from earlier versions of a system and applying them to later versions of the same system \cite{LTRWES, FARSEC, 9371393}. This approach is particularly effective when a substantial amount of historical data is available for a project. However, it is still difficult for engineers to use past information to accurately predict future bugs because, in reality, the proportion of SBRs in BRs is deficient, and future BRs may differ from those used for training.

Using the WPP method, we first utilize historical bug reports from each project to build prediction models and test them on future bug reports from the same project. 
To investigate the impact of data augmentation on predictor performance, we expand the training dataset with additional SBRs to assess how this increase affects prediction accuracy.
Next, we incorporate NSBRs from external datasets to evaluate their effect on SBR predictions. Our approach explores whether enriching the training dataset with both historical bug reports from the project and data from external projects can improve the model's accuracy and efficiency in predicting potential future SBRs. The final training dataset combines the WPP data with information from other projects.

\subsubsection{Cross-Project Prediction (CPP)}
When a project lacks BRs for model training, CPP can provide a solution by leveraging relevant data from other projects. CPP utilizes labeled BRs from external projects to predict unlabeled future BRs for a target project.
We aim to explore how information from other projects can enhance the prediction of SBRs in the target project by leveraging knowledge from these external sources. Additionally, we seek to evaluate the extent to which this integration influences the model’s ability to predict future SBRs from the excluded dataset.
During the training phase, we gradually integrate additional bug reports from other datasets, utilizing their complete training and testing splits. The latter half of one specific dataset is designated solely for testing. Using the testing splits from external projects ensures the availability of future bug reports, allowing us to assess how effectively the models can utilize this knowledge to predict future SBRs in the target project.

\section{Experimental Setup}
\label{sec:ExperimentalSetup}

This section describes the datasets used in this study, followed by three distinct evaluations of the model's performance in predicting future unseen SBRs. Finally, it introduces the performance metrics employed to assess the model's effectiveness.

\subsection{Dataset Setting}

Previous studies \cite{Ohiraetal, FARSEC, 9371393, LTRWES, CASMS} have extensively relied on five datasets, namely Chromium, Ambari, Wicket, Derby, and Camel, for security bug prediction. We use the refined version of these datasets \cite{9371393}, listed in Table \ref{tab:datasetsDetails}. Each row corresponds to a bug report, with columns showing its features. We collected the essential features of issue ID, description, summary, and security. We combined the description and summary features to consolidate information into a single text field.

\subsection{Within-Project Prediction (WPP)}

\begin{figure}[htbp]
    \centering
    \caption{Sequential Dataset Partitioning for WPP}
    \vspace{10pt} 
    \includegraphics[width=0.35\textwidth]{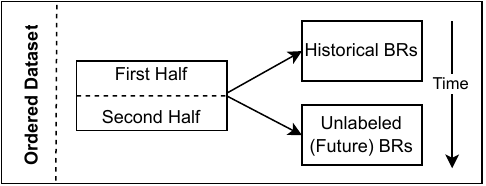}
    \label{fig:historicalBugReports}
\end{figure}

In WPP, models are trained and tested on a single dataset. As shown in Figure \ref{fig:historicalBugReports}, each dataset is ordered chronologically and then split into two halves (50\% each). The first half contains historical bug reports, representing earlier BRs used to train, while the second half consists of unlabeled future BRs, used to evaluate the model's performance.

\begin{table}[htpb]
\caption{Bug Report Distribution in WPP}
\label{tab:TreatmentsAndDatasets-WPP}
\footnotesize 
\setlength{\tabcolsep}{7pt}
\renewcommand{\arraystretch}{1.7}
\begin{tabular}{lcccl}
\hline
\multirow{2}{*}{\textbf{Target Dataset}} & \multicolumn{2}{c}{\textbf{Train Set}} & \multicolumn{2}{c}{\textbf{Test Set}} \\ \cline{2-3} \cline{4-5} 
 & \textbf{SBR (\%)} & \textbf{NSBR (\%)} & \textbf{SBR (\%)} & \textbf{NSBR (\%)} \\ \hline\hline
Chromium & 371 (46\%) & 20599 (50\%) & 437 (54\%) & 20533 (50\%) \\
Derby & 82 (46\%) & 418 (51\%) & 97 (54\%) & 403 (49\%) \\
Camel & 28 (38\%) & 472 (51\%) & 46 (62\%) & 454 (49\%) \\
Ambari & 40 (72\%) & 460 (49\%) & 16 (28\%) & 484 (51\%) \\
Wicket & 24 (52\%) & 476 (50\%) & 23 (48\%) & 477 (50\%) \\ \hline
\end{tabular}
\end{table}

The distribution of SBRs and NSBRs for the chronologically ordered datasets is shown in Table \ref{tab:TreatmentsAndDatasets-WPP}. During the training process, the training subset is further divided into 10\% for validation and 90\% training for parameter tuning. After completing the training, we assess the model's performance on the testing subset.

\begin{table}[htpb]
\caption{Bug Report Distribution in WPP using FARSEC}
\label{tab:FARSEC-WPP}
\footnotesize 
\setlength{\tabcolsep}{7pt}
\renewcommand{\arraystretch}{1.7}
\begin{tabular}{lclcl}
\hline
\multirow{2}{*}{\textbf{Target Dataset}} & \multicolumn{2}{c}{\textbf{Train Set}} & \multicolumn{2}{c}{\textbf{Test Set}} \\ \cline{2-3} \cline{4-5} 
 & \textbf{SBR (\%)} & \textbf{NSBR (\%)} & \textbf{SBR (\%)} & \textbf{NSBR (\%)} \\ \hline\hline
Chromium & 371 (45.9\%) & 20493 (49.9\%) & 437 (54.1\%) & 20533 (50.1\%) \\
Derby & 82 (45.8\%) & 46 (10.2\%) & 97 (54.2\%) & 403 (89.8\%) \\
Camel & 28 (37.8\%) & 214 (32\%) & 46 (62.2\%) & 454 (68\%) \\
Ambari & 40 (71.4\%) & 229 (32.1\%) & 16 (28.6\%) & 484 (67.9\%) \\
Wicket & 24 (51.1\%) & 207 (30.3\%) & 23 (48.9\%) & 477 (69.7\%) \\ \hline
\end{tabular}
\end{table}

Initially, we train BERT and RF models on five datasets. Subsequently, we apply the FARSEC filtering method to each dataset, effectively removing misleading NSBRs and generating new, refined datasets. In WPP, we develop a total of 20 prediction models: 10 models incorporate the FARSEC, and 10 use only BERT and RF.

Upon applying the ``farsectwo'' filtering method to the resulting datasets, the proportion of SBRs and NSBRs is presented in Table \ref{tab:FARSEC-WPP}.  In this Table, the count of NSBRs in the training datasets generally decreases across all target datasets compared to Table \ref{tab:TreatmentsAndDatasets-WPP}.

\subsubsection{Augmenting Data}

Given the low proportion of SBRs in a project, filtering NSBRs results in information loss. To address this, we explore the impact of incorporating data from other projects on SBR prediction performance while preserving all BRs. In this analysis, the training set comprises 50\% historical BRs from a specific project, enriched with data from other projects. The training dataset builds upon the WPP training set used in the WPP analysis, combined with parts of other datasets.

Algorithm \ref{Algorithm1} depicts our data augmentation approach by leveraging data from other projects. We iteratively select each dataset as a primary source, splitting it for training and testing. For every primary dataset, the training data is then augmented in two ways:
\begin{enumerate}
\item Gradually adding SBRs from other datasets at a time
\item Gradually adding all BRs from other datasets at a time
\end{enumerate}

Performance metrics such as recall, precision, F1-score, and G-measure are calculated to assess the impact of data augmentation, and results are collected for analysis.

\begin{algorithm}
\caption{Data Augmentation}
\begin{algorithmic}[1]
\label{Algorithm1}
\footnotesize
\setlength{\tabcolsep}{8pt}
\renewcommand{\arraystretch}{1.9}
\REQUIRE \texttt{Datasets[]} 
\ENSURE \texttt{PerformanceMetrics} 
\STATE \textbf{Initialization:} \texttt{Results[]} $\gets$ \texttt{[]}

\FOR{\texttt{type} in [\texttt{AllBRs}, \texttt{SBRs} ]}
    \FOR{\texttt{i} in \texttt{Datasets}}
        \STATE \texttt{Primary} $\gets$ \texttt{sortChron(Datasets[i])}
        \STATE Split \texttt{Primary} into \texttt{Train} and \texttt{Test}
        \STATE \texttt{AugmentedTrain} $\gets$ \texttt{Train}
        \FOR{\texttt{j} in \texttt{Datasets} - \texttt{i}}
            \STATE \texttt{BRs} $\gets$ \texttt{Datasets[j]\textsubscript{type}}
            \STATE \texttt{AugmentedTrain} $\gets$ \texttt{augmentDataset(AugmentedTrain, BRs)}
        \ENDFOR
        \STATE \texttt{Model} $\gets$ \texttt{trainModel(AugmentedTrain)}
        \STATE \texttt{Metrics} $\gets$ \texttt{evaluateModel(Model, Test)}
        \STATE \texttt{Results.append(Model\textsubscript{type}, Metrics)}
    \ENDFOR
\ENDFOR
\RETURN \texttt{Results}
\end{algorithmic}
\end{algorithm}

Table \ref{tab:TreatmentsAndDatasets-AugmentationBRs} shows the distribution of augmented datasets, including all BRs from additional datasets. 
For instance, in the first row, the training set includes the Chromium WPP training set augmented with all BRs from Derby, Camel, Wicket, and Ambari. The test set consists of the Chromium test set with 437 SBRs and 20,533 NSBRs, identical to the test set used in the WPP analysis.

\begin{table}[htpb]
\caption{Bug Report Distribution with Augmented SBRs and NSBRs}
\label{tab:TreatmentsAndDatasets-AugmentationBRs}
\footnotesize 
\setlength{\tabcolsep}{7pt}
\renewcommand{\arraystretch}{1.7}

\begin{tabular}{lclcl}
\hline
\multirow{2}{*}{\textbf{Target Dataset}} & \multicolumn{2}{c}{\textbf{Train Set}} & \multicolumn{2}{c}{\textbf{Test Set}} \\ \cline{2-3} \cline{4-5} 
 & \textbf{SBR (\%)} & \textbf{NSBR (\%)} & \textbf{SBR (\%)} & \textbf{NSBR (\%)} \\ \hline\hline
Chromium & 727 (62\%) & 24243 (54\%) & 437 (38\%) & 20533 (46\%) \\
Derby & 1067 (92\%) & 44373 (99\%) & 97 (8\%) & 403 (1\%) \\
Camel & 1118 (96\%) & 44322 (99\%) & 46 (4\%) & 454 (1\%) \\
Ambari & 1148 (99\%) & 44292 (99\%) & 16 (1\%) & 484 (1\%) \\
Wicket & 1141 (98\%) & 44299 (99\%) & 23 (2\%) & 477 (1\%) \\  \hline

\end{tabular}

\end{table}

Table \ref{tab:TreatmentsAndDatasets-AugmentationSBRs} presents the distribution of augmented datasets, incorporating all SBRs from additional datasets. For instance, in the first row, the training set includes the Chromium WPP training set with all SBRs from Derby, Camel, Wicket, and Ambari. The test set consists of the Chromium test set identical to the WPP analysis. 

\begin{table}[htpb]
\caption{Bug Report Distribution with Augmented SBRs}
\label{tab:TreatmentsAndDatasets-AugmentationSBRs}
\footnotesize 
\setlength{\tabcolsep}{7pt}
\renewcommand{\arraystretch}{1.7}

\begin{tabular}{lclcl}
\hline
\multirow{2}{*}{\textbf{Target Dataset}} & \multicolumn{2}{c}{\textbf{Train Set}} & \multicolumn{2}{c}{\textbf{Test Set}} \\ \cline{2-3} \cline{4-5} 
 & \textbf{SBR (\%)} & \textbf{NSBR (\%)} & \textbf{SBR (\%)} & \textbf{NSBR (\%)} \\ \hline\hline
Chromium & 727 (62\%) & 20599 (50\%) & 437 (38\%) & 20533 (50\%) \\
Derby &  1067 (92\%) &  418 (51\%) & 97 (8\%) & 403 (49\%) \\
Camel &  1118 (96\%) & 472 (51\%) & 46 (4\%) & 454 (49\%) \\
Ambari &  1148 (99\%) & 460 (49\%) & 16 (1\%) & 484 (51\%) \\
Wicket & 1165 (98\%) & 476 (50\%) & 23 (2\%) & 477 (50\%) \\  \hline
\end{tabular}
\end{table}

Due to space constraints, we focus on the results of adding all BRs and SBRs, with full data available in our replication package. We will discuss the outcomes of incrementally adding information (SBRs and BRs) further in our discussion. 
Using five datasets and two learning approaches, we can create a total of 150 models for BERT and RF using additional BRs. Adding one dataset results in \(5 \times 4 \times 2 learners = 40\) models. Adding two extra datasets results in \(5 \times \binom{4}{2} \times 2 learners= 60\) models. Adding three extra datasets leads to \(5 \times \binom{4}{3} \times 2 learners = 40\) models, and finally, there are 10 models when including all four additional datasets. In total, this sums to \(40 + 60 + 40 + 10 = 150\) possible models. This approach is consistently applied for both BRs and SBRs, yielding 2×150=300 models.

\subsection{Cross-Project Prediction (CPP)}

CPP leverages labeled BRs from one or more projects to predict unlabeled future bug reports for a target project. We incrementally include additional BRs from other datasets during the training phase, utilizing their full training and testing splits. The latter half of one specific dataset is used exclusively for testing.

In CPP, we follow the same method of incrementally adding datasets. First, we select one dataset as the target, using only its test set for evaluation. Next, we select a single external dataset, utilizing all its BRs for training, and then iteratively incorporate additional datasets into the training set. At each step, we calculate performance metrics to assess the impact of this method. The difference between CPP and augmentation lies in how the BRs of the target dataset are utilized. In augmentation, the historical BRs from the training set are used for training purposes. In contrast, in CPP, the training set is excluded from the training, and the test set is solely employed for prediction.

Table \ref{tab:TreatmentsAndDatasets-CPP} presents the distribution of BRs in the CPP datasets when incorporating all four datasets for training. As an example, the first analysis of this table includes complete datasets from Derby, Camel, Ambari, and Wicket, totaling 356 SBRs and 3,644 NSBRs while excluding the Chromium dataset from training. The second half of the ordered Chromium dataset, containing 437 SBRs and 20,533 NSBRs, is used as the testing set. Due to space constraints, we do not detail the bug distribution in CPP at every step, as can be inferred from Table \ref{tab:datasetsDetails}. 

We constructed a total of \( \binom{5}{1} \times 2 \) + \( \binom{5}{2} \times 2 \) + \( \binom{5}{3} \times 2\) + \( \binom{5}{1} \times 2\)=
 (10 + 20 + 20 + 10) = 60 prediction models by utilizing five datasets and two learning algorithms.

\begin{table}[htpb]
\caption{Bug Report Distribution for CPP Using All Datasets}
\label{tab:TreatmentsAndDatasets-CPP}
\footnotesize
\setlength{\tabcolsep}{7pt}
\renewcommand{\arraystretch}{1.7}
\begin{tabular}{lclcl}
\hline
\multirow{2}{*}{\textbf{Target Dataset}} & \multicolumn{2}{c}{\textbf{Train Set}} & \multicolumn{2}{c}{\textbf{Test Set}} \\ \cline{2-3} \cline{4-5} 
 & \textbf{SBR (\%)} & \textbf{NSBR (\%)} & \textbf{SBR (\%)} & \textbf{NSBR (\%)} \\ \hline\hline
Chromium & 356 (45\%) & 3644 (15\%) & 437 (55\%) & 20533 (85\%) \\
Derby & 985 (91\%) & 43955 (99\%) & 97 (9\%) & 403 (1\%) \\
Camel & 1090 (96\%) & 43850 (99\%) & 46 (4\%) & 454 (1\%) \\
Ambari & 1108 (99\%) & 43832 (99\%) & 16 (1\%) & 484 (1\%) \\ 
Wicket & 1117 (98\%) & 43823 (99\%) & 23 (2\%) & 477 (1\%) \\ \hline
\end{tabular}
\end{table}

\subsection{Performance Metrics}

For each bug report, the prediction result can yield four possible outcomes, as detailed in Table \ref{tab:ConfusionMatrix-table}, from which the performance metrics are derived.

\begin{table}[htpb]
\centering
\caption{Confusion Matrix}
\label{tab:ConfusionMatrix-table}
\resizebox{0.95\columnwidth}{!}{%
\begin{tabular}{lp{0.2\columnwidth}p{0.2\columnwidth}p{0.2\columnwidth}} 
 & \multicolumn{1}{c}{}  & \multicolumn{2}{c}{Predict}                             \\ \cline{3-4} 
 & \multicolumn{1}{l|}{} & \multicolumn{1}{l|}{SBRs} & \multicolumn{1}{l|}{NSBRs} \\ \cline{2-4} 
\multicolumn{1}{l|}{\multirow{2}{*}{Actual}} & \multicolumn{1}{l|}{SBRs}  & \multicolumn{1}{l|}{True Positive (TP)} & \multicolumn{1}{l|}{False Negative (FN)} \\ \cline{2-4} 
\multicolumn{1}{l|}{}                         & \multicolumn{1}{l|}{NSBRs} & \multicolumn{1}{l|}{False Positive (FP)} & \multicolumn{1}{l|}{True Negative (TN)} \\ \cline{2-4} 
\end{tabular}%
}
\end{table}

The evaluation metrics used in this study are as follows: \textit{Precision} measures the fraction of actual SBRs among the predicted SBRs, reflecting the accuracy of positive predictions as defined in Equation \ref{eq:precision}. \textit{Recall} assesses the proportion of correctly classified SBRs among all verified SBRs, indicating the probability of detection, as shown in Equation \ref{eq:recall}. The \textit{F1-score} is the harmonic mean of \textit{Precision} and \textit{Recall}, providing a balance between these two metrics to evaluate overall performance, as detailed in Equation \ref{eq:f1score}. The \textit{Probability of False Alarm (FPR)} quantifies the fraction of NSBRs mistakenly identified as SBRs, representing the false positive rate, as expressed in Equation \ref{eq:pf}. Finally, the \textit{G-measure} combines \textit{Recall} with the complement of FPR, offering a balanced evaluation of a model’s performance by capturing its effectiveness in identifying true positives while minimizing false alarms, as indicated in Equation \ref{eq:gmeasure}.

\begin{center}
\small 
\begin{align*}
    \text{Precision} &= \frac{TP}{TP + FP} \tag{1} \label{eq:precision}
\end{align*}
\end{center}

\begin{center}
\small 
\begin{align*}
    \text{Recall} &= \frac{TP}{TP + FN} \tag{2} \label{eq:recall}
\end{align*}
\end{center}

\begin{center}
\small 
\begin{align*}  
    \text{F1-score} &= \frac{2 \times \text{Recall} \times \text{Precision}}{\text{Recall} + \text{Precision}} \tag{3} \label{eq:f1score}
\end{align*}
\end{center}

\begin{center}
\small 
\begin{align*}  
   pf = FPR &= \frac{FP}{FP + TN} \tag{4} \label{eq:pf}
\end{align*}
\end{center}

\begin{center}
\small 
\begin{align*}  
   \text{G-measure} &= \frac{2 \times \text{Recall} \times (1 - pf)}{\text{Recall} + (1 - pf)} \tag{5} \label{eq:gmeasure}
\end{align*}
\end{center}

\vspace{1\baselineskip}
All these metrics range from 0 to 1. Among the five performance metrics, \textit{Recall}, \textit{Precision}, \textit{F1-score}, and \textit{G-measure} are preferred to be higher, indicating better performance, while the \textit{FPR} is preferred to be lower.

\section{EXPERIMENT RESULTS}
\label{sec:EXPERIMENTRESULTS}

In the WPP section, we will address RQ1 and RQ2 in the initial part, while RQ3 will be covered in the CPP section.
Finally, in the last section, we will compare the results of the WPP and CPP analyses.

\subsection{Within Project Results}

\begin{tcolorbox}[RQsboxstyle]
    \textbf{RQ1:} How effectively can BERT predict SBRs compared to the state-of-the-art RF?
\end{tcolorbox}

\begin{table}[htpb]
\caption{BERT versus RF in WPP}
\label{tab:RQ1}
\renewcommand{\arraystretch}{1.7} 
\setlength{\tabcolsep}{2.5pt} 
\centering 
\footnotesize
\begin{tabular}{lcccccccccc}
\hline
\multirow{2}{*}{Dataset} & \multicolumn{2}{c}{Recall} & \multicolumn{2}{c}{Precision} & \multicolumn{2}{c}{F1-score} & \multicolumn{2}{c}{FPR} & \multicolumn{2}{c}{G-Measure} \\ \cline{2-11} 
 & BERT & RF & BERT & RF & BERT & RF & BERT & RF & BERT & RF \\ \hline \hline
Chromium & 0.70 & 0.61 & 0.90 & 0.96 & 0.79 & 0.74 & 0.00 & 0.00 & 0.83 & 0.75 \\ \hline
Derby & 0.40 & 0.45 & 0.68 & 0.94 & 0.51 & 0.61 & 0.04 & 0.01 & 0.57 & 0.62 \\ \hline
Camel & 0.20 & 0.33 & 0.60 & 0.94 & 0.30 & 0.48 & 0.01 & 0.00 & 0.33 & 0.49 \\ \hline
Ambari & 0.25 & 0.44 & 0.31 & 0.37 & 0.28 & 0.40 & 0.02 & 0.02 & 0.40 & 0.60 \\ \hline
Wicket & 0.13 & 0.52 & 0.43 & 0.80 & 0.20 & 0.63 & 0.01 & 0.01 & 0.23 & 0.68 \\ \hline
\textbf{Average} & \textbf{0.34} & \textbf{0.47} & \textbf{0.58} & \textbf{0.80} & \textbf{0.41} & \textbf{0.57} & \textbf{0.02} & \textbf{0.01} & \textbf{0.47} & \textbf{0.63} \\ \hline
\end{tabular}
\end{table}

Table \ref{tab:RQ1} presents the performance of BERT and RF in predicting SBRs within the same datasets. The results demonstrate that RF outperforms BERT in all metrics on average of the five datasets.
RF achieves higher performance than BERT in recall, precision, F1-score, and G-measure, while showing a lower FPR. Note that a lower FPR is preferable.

In a real application scenario where software vendors monitor emerging bug reports related to their products, recall is crucial, as missing threats could result in significant losses \cite{10.1145/3540250.3549156, 10.1007/s10515-014-0162-2}. We prioritize the G-measure, which combines recall and the FPR, as a key evaluation metric. This allows us to assess the efficacy of our models in predicting SBRs while minimizing unnecessary actions caused by misclassification of NSBRs as SBRs.

\begin{tcolorbox}[yellowboxstyle]
    The results show that RF outperforms BERT with a 34\% higher average G-measure for predicting SBRs in WPP.
\end{tcolorbox}

A comparison of BERT and RF for each dataset reveals that Wicket showed the most improvement in G-Measure, followed by Ambari and Camel, while Derby showed the least improvement. In contrast, when trained on Chromium, BERT outperforms RF in G-Measure, with a similar trend observed in recall. However, RF exhibits slightly better precision, leading to a more balanced F1-score.

Table \ref{tab:TreatmentsAndDatasets-WPP} outlines the percentage distribution of SBRs and NSBRs, highlighting security vulnerabilities and non-security bugs across systems. Chromium and Derby maintain a balanced distribution between training and testing datasets. For instance, Chromium had 371 SBRs and 20,599 NSBRs for training, and 437 SBRs and 20,533 NSBRs for testing. In contrast, Camel, Ambari, and Wicket exhibit imbalances. For example, Ambari had 40 SBRs for training and only 16 SBRs for testing. This variation impacts BERT's performance. In contrast, this disparity appears less influential for RF, as the recall values across all five datasets remain within a consistent range.

\begin{tcolorbox}[yellowboxstyle]
    While RF outperforms BERT in average G-measure in WPP, their effectiveness varies across different datasets.
\end{tcolorbox}

\begin{table}[htpb]
\caption{BERT versus RF in WPP using FARSEC }
\label{tab:RQ11}
\centering 
\renewcommand{\arraystretch}{1.7} 
\setlength{\tabcolsep}{2.5pt} 
\footnotesize
\begin{tabular}{lccccccccccc}
\hline
\multirow{2}{*}{Dataset} & \multicolumn{2}{c}{Recall} & \multicolumn{2}{c}{Precision} & \multicolumn{2}{c}{F1-score} & \multicolumn{2}{c}{FPR} & \multicolumn{2}{c}{G-Measure} \\ \cline{2-11} 
 & BERT & RF & BERT & RF & BERT & RF & BERT & RF & BERT & RF \\ \hline \hline
Chromium & 0.71 & 0.61 & 0.92 & 0.94 & 0.80 & 0.74 & 0.00 & 0.00 & 0.83 & 0.76 \\ \hline
Derby & 0.85 & 0.76 & 0.24 & 0.27 & 0.38 & 0.40 & 0.63 & 0.49 & 0.52 & 0.61 \\ \hline
Camel & 0.24 & 0.39 & 0.21 & 0.90 & 0.22 & 0.55 & 0.09 & 0.00 & 0.38 & 0.56 \\ \hline
Ambari & 0.50 & 0.69 & 0.10 & 0.44 & 0.16 & 0.54 & 0.16 & 0.03 & 0.63 & 0.81 \\ \hline
Wicket & 0.22 & 0.60 & 0.23 & 0.60 & 0.22 & 0.60 & 0.04 & 0.01 & 0.35 & 0.75 \\ \hline
\textbf{Average} & \textbf{0.50} & \textbf{0.61} & \textbf{0.34} & \textbf{0.63} & \textbf{0.36} & \textbf{0.57} & \textbf{0.18} & \textbf{0.11} & \textbf{0.54} & \textbf{0.70} \\ \hline
\end{tabular}
\end{table}

By applying the FARSEC algorithm to compute TF-IDF values for terms in SBRs, we identified the top 100 terms as security-related keywords. Filtering NSBRs using these keywords created new training datasets based on FARSEC. Table \ref{tab:RQ11} presents the performance results of BERT and RF evaluated in these datasets. The comparison of BERT and RF performance metrics for predicting SBRs using FARSEC reveals patterns similar to those observed in the WPP analysis, indicating that RF outperforms BERT with an average G-measure of 0.70, compared to BERT's 0.54. 
Across datasets, Wicket shows the most improvement, while Derby shows the least. However, in the Chromium dataset, BERT outperforms RF, achieving a G-measure of 0.83 versus RF's 0.76.

\begin{tcolorbox}[yellowboxstyle]
    Applying FARSEC both BERT and RF exhibit improved average G-measure performance, with RF often outperforming BERT across all datasets.
\end{tcolorbox}

Comparing Tables \ref{tab:RQ1} and \ref{tab:RQ11}, we observed that for the Chromium dataset, the G-Measure using BERT and RF remains nearly unchanged. This indicates that the FARSEC filtering approach does not significantly impact the prediction of SBRs in this dataset with the largest proportion of BRs. The application of FARSEC decreased NSBRs in the datasets as expected, as shown in Table \ref{tab:FARSEC-WPP}. Notably, the Derby dataset experienced a significant reduction in NSBRs, from 418 to 46, while the number of SBRs remains constant. This imbalance during training biases the model toward identifying more SBRs, resulting in a higher rate of false positives, with BERT reaching 0.63 and RF reaching 0.49.

Aside from the Chromium and Derby datasets, the FARSEC filtering approach generally improves G-measure using RF and BERT. For instance, in the Ambari dataset, the G-measure achieved in WPP with BERT and RF are 0.40 and 0.60, respectively. After applying FARSEC, these measures improve to 0.63 for BERT and 0.81 for RF. However, in the Chromium and Derby datasets, the changes were negligible.

In our study, we tested the RF model in combination with SMOTE, as explored in Wu et al.'s research. Despite this, the results averaged across all five datasets indicated that the simple RF model outperformed the upsampled variant. This finding does not entirely align with the results of Wu et al., which showed that simple text classification models surpass both FARSEC and its various tuning methods \cite{9371393}.

\begin{tcolorbox}[yellowboxstyle]
    Except for the Chromium and Derby datasets, FARSEC improves the G-measure using RF and BERT, differing from Wu et al., who found simple text classification models outperforming FARSEC.
\end{tcolorbox}

\begin{tcolorbox}[RQsboxstyle]
    \textbf{RQ2:} What is the impact of augmenting data on the performance of security bug report predictors?
\end{tcolorbox}

In this research, we evaluate BERT and RF's prediction performance by augmenting BRs from other projects. Initially, we added SBRs from other datasets. Then, we assessed the impact of including all BRs, such as NSBRs, in the training set, reflecting real-world scenarios where NSBRs usually outnumber SBRs.

We evaluated our models by comparing them with WPP models, using consistent testing datasets across all experiments. Given the time complexity of FARSEC, O(N×W), where W represents the number of security-related keywords in the feature set and N denotes the number of BRs in the training data, and the fact that it removes SBR-like information from NSBRs, we aimed to retain NSBR information in our analysis. Therefore, we focused on RF and BERT as baseline classifiers without applying FARSEC to the training datasets.

\begin{table}[htpb]
\caption{BERT versus RF using Augmented Datasets}
\label{tab:RQ2BRs}
\renewcommand{\arraystretch}{1.7} 
\setlength{\tabcolsep}{2.5pt}
\centering 
\footnotesize
\begin{tabular}{lcccccccccc}
\hline
\multirow{2}{*}{Dataset} & \multicolumn{2}{c}{Recall} & \multicolumn{2}{c}{Precision} & \multicolumn{2}{c}{F1-score} & \multicolumn{2}{c}{FPR} & \multicolumn{2}{c}{G-Measure} \\ \cline{2-11} 
 & BERT & RF & BERT & RF & BERT & RF & BERT & RF & BERT & RF \\ \hline \hline
Chromium \scriptsize{\textsubscript{SBRs}} & 0.72 & 0.60 & 0.90 & 0.95 & 0.80 & 0.74 & 0.00 & 0.00 & 0.84 & 0.75 \\
\multicolumn{1}{l}{Chromium \scriptsize{\textsubscript{BRs}}} & 0.73 & 0.62 & 0.88 & 0.95 & 0.80 & 0.75 & 0.00 & 0.00 & 0.84 & 0.77 \\ \hline
Derby \scriptsize{\textsubscript{SBRs}} & 0.59 & 0.46 & 0.67 & 0.80 & 0.63 & 0.58 & 0.07 & 0.03 & 0.72 & 0.62 \\
Derby \scriptsize{\textsubscript{BRs}} & 0.59 & 0.18 & 0.77 & 0.54 & 0.67 & 0.27 & 0.04 & 0.03 & 0.73 & 0.30 \\ \hline
Camel \scriptsize{\textsubscript{SBRs}} & 0.28 & 0.45 & 0.37 & 0.60 & 0.32 & 0.51 & 0.05 & 0.03 & 0.43 & 0.61 \\
Camel \scriptsize{\textsubscript{BRs}} & 0.41 & 0.26 & 0.73 & 0.92 & 0.53 & 0.40 & 0.01 & 0.00 & 0.58 & 0.41 \\ \hline
Ambari \scriptsize{\textsubscript{SBRs}} & 0.31 & 0.68 & 0.11 & 0.23 & 0.17 & 0.35 & 0.08 & 0.07 & 0.46 & 0.78 \\
Ambari \scriptsize{\textsubscript{BRs}} & 0.19 & 0.37 & 0.38 & 0.37 & 0.25 & 0.37 & 0.01 & 0.02 & 0.32 & 0.54 \\ \hline
Wicket \scriptsize{\textsubscript{SBRs}} & 0.35 & 0.56 & 0.33 & 0.61 & 0.34 & 0.59 & 0.03 & 0.02 & 0.51 & 0.71 \\
Wicket \scriptsize{\textsubscript{BRs}} & 0.70 & 0.17 & 0.84 & 0.67 & 0.76 & 0.28 & 0.01 & 0.00 & 0.82 & 0.29 \\ \hline
\textbf{Average \scriptsize{\textsubscript{SBRs}}} & \textbf{0.45} & \textbf{0.55} & \textbf{0.48} & \textbf{0.64} & \textbf{0.45} & \textbf{0.55} & \textbf{0.05} & \textbf{0.03} & \textbf{0.59} & \textbf{0.70} \\
\textbf{Average \scriptsize{\textsubscript{BRs}}} & \textbf{0.52} & \textbf{0.32} & \textbf{0.72} & \textbf{0.69} & \textbf{0.60} & \textbf{0.41} & \textbf{0.02} & \textbf{0.01} & \textbf{0.66} & \textbf{0.46} \\ \hline
\end{tabular}
\end{table}

Table \ref{tab:RQ2BRs} presents the performance of the BERT and RF classifiers on the augmented datasets. Subscripts SBRs and BRs denote the augmented data type. As an example, Chromium{\scriptsize{\textsubscript{SBRs}}} is Chromium augmented with SBRs from other projects, while Chromium{\scriptsize{\textsubscript{BRs}}} is augmented with BRs.

The results indicate that using only SBRs from other datasets often enables RF to outperform BERT in recall, F1-score, and G-measure of each dataset. However, this is not the case for the Chromium and Derby datasets, where BERT excels in G-measure performance.

\begin{tcolorbox}[yellowboxstyle]
    Augmenting datasets with more SBRs often results in RF outperforming BERT in terms of G-measure across all datasets, except Chromium and Derby.
\end{tcolorbox}

Table \ref{tab:RQ2BRs} shows that in augmented datasets containing all BRs, BERT outperforms RF in G-measure and F1-score across all datasets except Ambari, with notable improvements in Wicket, Derby, and Camel. For example, in the Derby dataset, BERT achieves a G-measure of 0.73, compared to RF’s 0.30. Ambari, however, exhibits lower performance for BERT, likely due to having the fewest SBRs in the test set (only 16), leading to a distribution imbalance that differs from other datasets, as shown in Table \ref{tab:TreatmentsAndDatasets-AugmentationBRs}.

Comparing the impact of adding SBRs to that of adding BRs, in the Derby dataset, when augmented with SBRs, BERT and RF have G-measures of 0.72 and 0.62, respectively. When BRs are included, BERT achieves a G-measure of 0.73, while RF drops to 0.30. Similarly, in the Wicket dataset, BERT and RF start with G-Measures of 0.51 and 0.71 with SBRs, but with all BRs, BERT improves to 0.82, whereas RF decreases to 0.29.
In the Ambari dataset, both BERT and RF experience a decline in G-measure. In contrast, in the Chromium dataset, RF shows a small improvement with bug report augmentation, while BERT's performance remains stable.
According to this table, RF achieves a best G-measure of 0.70, surpassing BERT's 0.66 across all datasets.

\begin{tcolorbox}[yellowboxstyle]
    Adding only SBRs enhances the average performance of both models. However, when both SBRs and NSBRs are included, RF's average performance significantly drops to 0.46, while BERT achieves its highest average performance of 0.66, surpassing RF.
\end{tcolorbox}

Comparing tables \ref{tab:RQ2BRs} and \ref{tab:RQ1}, adding more SBRs improves the G-Measure for both BERT and RF in the Camel, Ambari, Derby, and Wicket datasets. 
Among these, BERT shows a more significant improvement than RF in the Wicket and Derby datasets, suggesting that adding SBRs is particularly advantageous. In the Derby dataset, for instance, BERT achieves a 0.15 increase, while RF remains unchanged. Similarly, in the Wicket dataset, BERT shows a 0.23 increase compared to RF's 0.03. However, in the Ambari dataset, BERT experiences a decrease in both F1-score and precision, while the Wicket dataset sees an increase in F1-score from 0.20 to 0.34. In contrast, in the Chromium dataset, which initially had more SBRs, BERT surpasses RF. In Chromium, RF's G-measure remains constant, while BERT sees a 0.01 increase. This may suggest that, with an adequate amount of SBRs, further SBRs may not offer additional benefits.

Comparing Tables \ref{tab:RQ2BRs} and \ref{tab:RQ1}, applying BERT to datasets augmented with BRs significantly improves prediction performance over WPP, boosting G-measure and F1-score in most cases except for Ambari. The improvement varies by dataset—for instance, in Wicket, BERT’s G-measure jumps from 0.23 to 0.82, while in Derby, it increases more modestly from 0.57 to 0.73. Conversely, Ambari sees a decline, with G-measure dropping from 0.40 to 0.32.
RF, however, generally performs worse with BR-augmented datasets, except for Chromium, where its G-measure improves slightly from 0.75 to 0.77. The decline is more substantial in Derby and Wicket compared to Camel and Ambari.

\subsection{Cross-Project Results}

\begin{tcolorbox}[RQsboxstyle]
    \textbf{RQ3:} How do BERT and RF models compare for cross-project prediction?
\end{tcolorbox}

To tackle the challenge of underrepresented SBRs, we assessed prediction performance by first augmenting SBRs from other projects and subsequently adding BRs, as outlined in RQ2. This approach was particularly effective when substantial historical data is available for a project. 
However, when BRs are limited or unavailable, one potential solution is to leverage external data from other projects for training. This research question explores whether models trained on BRs from one or more projects can generalize to accurately detect SBRs in unseen projects.

\begin{table}[htpb] 
\caption{BERT versus RF in CPP using a Single External Dataset} 
\label{tab:RQ32} 
\renewcommand{\arraystretch}{1.7} 
\setlength{\tabcolsep}{2.5pt} 
\centering 
\footnotesize
\begin{tabular}{lcccccccccc}
\hline
\multicolumn{1}{c}{Dataset} & \multicolumn{2}{c}{Recall} & \multicolumn{2}{c}{Precision} & \multicolumn{2}{c}{F1-score} & \multicolumn{2}{c}{FPR} & \multicolumn{2}{c}{G-Measure} \\ \cline{2-11} 
\multicolumn{1}{c}{} & BERT & RF & BERT & RF & BERT & RF & BERT & RF & BERT & RF \\ \hline \hline
Chromium \scriptsize{\textsubscript{Derby}} & 0.27 & 0.32 & 0.06 & 0.22 & 0.10 & 0.26 & 0.09 & 0.02 & 0.42 & 0.48 \\
Chromium \scriptsize{\textsubscript{Camel}} & 0.12 & 0.02 & 0.20 & 0.02 & 0.15 & 0.02 & 0.01 & 0.02 & 0.21 & 0.04 \\
Chromium \scriptsize{\textsubscript{Ambari}} & 0.20 & 0.20 & 0.06 & 0.28 & 0.09 & 0.23 & 0.07 & 0.01 & 0.33 & 0.33 \\
Chromium \scriptsize{\textsubscript{Wicket}} & 0.44 & 0.32 & 0.84 & 0.36 & 0.57 & 0.34 & 0.00 & 0.01 & \textbf{0.61} & 0.48 \\ \hline

Derby \scriptsize{\textsubscript{Chromium}} & 0.07 & 0.03 & 0.44 & 0.75 & 0.12 & 0.05 & 0.02 & 0.00 & 0.13 & 0.06 \\
Derby \scriptsize{\textsubscript{Camel}} & 0.20 & 0.35 & 0.68 & 1.00 & 0.30 & 0.51 & 0.02 & 0.00 & 0.33 & \textbf{0.52} \\
Derby \scriptsize{\textsubscript{Ambari}} & 0.07 & 0.10 & 0.50 & 0.66 & 0.13 & 0.17 & 0.02 & 0.01 & 0.13 & 0.18 \\
Derby \scriptsize{\textsubscript{Wicket}} & 0.32 & 0.45 & 0.79 & 0.72 & 0.46 & 0.55 & 0.02 & 0.04 & 0.48 & 0.61 \\ \hline

Camel \scriptsize{\textsubscript{Chromium}} & 0.20 & 0.19 & 0.82 & 1.00 & 0.32 & 0.32 & 0.00 & 0.00 & 0.33 & 0.32 \\
Camel \scriptsize{\textsubscript{Derby}} & 0.28 & 0.41 & 0.65 & 0.86 & 0.39 & 0.55 & 0.02 & 0.01 & 0.44 & 0.58 \\
Camel \scriptsize{\textsubscript{Ambari}} & 0.07 & 0.04 & 0.30 & 0.50 & 0.11 & 0.08 & 0.02 & 0.00 & 0.13 & 0.08 \\
Camel \scriptsize{\textsubscript{Wicket}} & 0.26 & 0.56 & 0.92 & 0.89 & 0.41 & 0.69 & 0.00 & 0.01 & 0.41 & \textbf{0.72} \\ \hline

Ambari \scriptsize{\textsubscript{Chromium}} & 0.00 & 0.00 & 0.00 & 0.00 & 0.00 & 0.00 & 0.00 & 0.00 & 0.00 & 0.00 \\
Ambari \scriptsize{\textsubscript{Derby}} & 0.06 & 0.62 & 0.12 & 0.43 & 0.08 & 0.51 & 0.01 & 0.03 & 0.11 & \textbf{0.76} \\
Ambari \scriptsize{\textsubscript{Camel}} & 0.06 & 0.00 & 0.14 & 0.00 & 0.09 & 0.00 & 0.01 & 0.00 & 0.11 & 0.00 \\
Ambari \scriptsize{\textsubscript{Wicket}} & 0.06 & 0.06 & 0.50 & 0.33 & 0.11 & 0.10 & 0.00 & 0.00 & 0.11 & 0.11 \\ \hline

Wicket \scriptsize{\textsubscript{Chromium}} & 0.17 & 0.13 & 0.80 & 1.00 & 0.29 & 0.23 & 0.00 & 0.00 & 0.29 & 0.23 \\
Wicket \scriptsize{\textsubscript{Derby}} & 0.57 & 0.52 & 0.57 & 1.00 & 0.57 & 0.68 & 0.02 & 0.00 & \textbf{0.72} & 0.68 \\
Wicket \scriptsize{\textsubscript{Camel}} & 0.13 & 0.52 & 0.43 & 0.92 & 0.20 & 0.66 & 0.01 & 0.00 & 0.23 & 0.68 \\
Wicket \scriptsize{\textsubscript{Ambari}} & 0.09 & 0.08 & 0.12 & 1.00 & 0.10 & 0.16 & 0.03 & 0.00 & 0.16 & 0.15 \\ \hline
\multicolumn{1}{l}{\textbf{Average}} & \multicolumn{1}{l}{\textbf{0.18}} & \multicolumn{1}{l}{\textbf{0.25}} & \multicolumn{1}{l}{\textbf{0.45}} & \multicolumn{1}{l}{\textbf{0.60}} & \multicolumn{1}{l}{\textbf{0.23}} & \multicolumn{1}{l}{\textbf{0.31}} & \multicolumn{1}{l}{\textbf{0.02}} & \multicolumn{1}{l}{\textbf{0.01}} & \multicolumn{1}{l}{\textbf{0.29}} & \multicolumn{1}{l}{\textbf{0.35}}  \\ \hline
\end{tabular}
\end{table}

To assess whether bug reports from one project can effectively identify SBRs in another, we trained BERT and RF models on the full bug report set of a single dataset. We then evaluated them on the test sets of four other datasets. Table \ref{tab:RQ32} presents the performance of both models across different datasets.
The subscripts of the dataset names indicate the training datasets. For instance, Chromium{\scriptsize{\textsubscript{Derby}}} represents the prediction of SBRs in the Chromium test set using a model trained on all the BRs in the Derby dataset. In this table, the highest G-measures are bold.

Based on the average G-measure of all datasets, RF outperforms BERT with a G-measure of 0.35 compared to BERT’s 0.29. However, a closer look at the results for each dataset reveals no clear preference for either model. In some datasets, BERT performs better, while in others, RF is superior. For example, BERT outperforms RF on Chromium and Wicket, while RF is superior on Derby, Camel, and Ambari. 
Notably, the results indicate that Ambari exhibits the weakest predictive performance for both BERT and RF, except when training with the Derby dataset using RF, which boosts the G-measure to 0.76.
Overall, While there is a superior model, either RF or BERT, for predicting SBRs in each dataset, a model developed for one project may not yield effective results in another.

\begin{tcolorbox}[yellowboxstyle]
    Building a model based on one project may not effectively predict SBRs in another.
\end{tcolorbox}

\begin{table}[htpb] 

\caption{BERT versus RF in CPP using Multiple External Datasets} 
\label{tab:RQ3} 
\renewcommand{\arraystretch}{1.7}  
\setlength{\tabcolsep}{2.5pt}  
\centering 
\footnotesize
\begin{tabular}{lcccccccccc}
\hline
\multirow{2}{*}{Dataset} & \multicolumn{2}{c}{Recall} & \multicolumn{2}{c}{Precision} & \multicolumn{2}{c}{F1-score} & \multicolumn{2}{c}{FPR} & \multicolumn{2}{c}{G-Measure} \\ \cline{2-11} 
 & BERT & RF & BERT & RF & BERT & RF & BERT & RF & BERT & RF \\ \hline \hline
Chromium & 0.68 & 0.46 & 0.32 & 0.30 & 0.43 & 0.36 & 0.03 & 0.02 & 0.80 & 0.63 \\ \hline
Derby & 0.44 & 0.15 & 0.90 & 0.54 & 0.59 & 0.24 & 0.01 & 0.03 & 0.61 & 0.27 \\ \hline
Camel & 0.43 & 0.28 & 0.83 & 0.93 & 0.57 & 0.43 & 0.01 & 0.00 & 0.60 & 0.44 \\ \hline
Ambari & 0.13 & 0.06 & 0.50 & 0.50 & 0.20 & 0.11 & 0.00 & 0.00 & 0.22 & 0.12 \\ \hline
Wicket & 0.74 & 0.30 & 0.57 & 0.70 & 0.22 & 0.60 & 0.04 & 0.01 & 0.84 & 0.47 \\ \hline
\textbf{Average} & \textbf{0.48} & \textbf{0.25} & \textbf{0.62} & \textbf{0.59} & \textbf{0.49} & \textbf{0.31} & \textbf{0.02} & \textbf{0.01} & \textbf{0.62} & \textbf{0.38} \\ \hline
\end{tabular}
\end{table}

Table \ref{tab:RQ3} indicates that, in CPP, BERT achieves higher average F1-score and G-measure values compared to RF when using all the BRs from other datasets for training.

On average across all datasets, BERT achieves a G-Measure of 0.62, which is approximately 63\% higher than RF's 0.38. In the Chromium dataset, BERT achieves a G-measure of 0.80, while RF scores 0.63. Similarly, in the Wicket dataset, BERT attains a G-measure of 0.84, significantly outperforming RF's 0.47. These results indicate that BERT is more effective at predicting SBRs by utilizing all BRs from other datasets. The CPP analysis further reveals that Ambari displays the weakest predictive performance.

\begin{tcolorbox}[yellowboxstyle]
    Utilizing BRs from all other datasets for training, BERT achieves an average G-measure of 0.62, which is substantially higher than that of RF’s 0.38.
\end{tcolorbox}

While addressing RQ3, we found that although the average performance metrics indicated that using a single dataset for training and testing gives RF an advantage over BERT, the results became approximately equivalent after the addition of a second dataset. Subsequently, upon consolidating the third dataset, BERT demonstrated superior performance in predicting SBRs in an external dataset.

\subsection{Cross-Project vs. Within-Project}

In terms of the G-measures for each dataset, the WPP model outperforms the CPP in the Chromium dataset using both models. However, for datasets with fewer SBRs, such as Camel, Ambari, and Wicket, the top-performing model in the CPP surpasses the WPP results. For instance, the best predictor in the WPP for the Camel and Wicket datasets is RF, achieving G-measures of 0.49 and 0.68, respectively. In contrast, using BERT, we observed G-measures of 0.60 and 0.84, with the latter being the highest G-measure value recorded for the Wicket dataset.

Nonetheless, selecting one optimal dataset for training a model to predict SBRs in another is not feasible, as we cannot determine which dataset is best suited. As an example, Wicket is most effective for Chromium, while Derby excels with Ambari and Wicket.
Our investigation suggests that relying solely on single sources for building SBR predictors is not advisable. This contrasts with Peters et al. \cite{FARSEC}, who found that transfer project prediction models can significantly enhance performance, outperforming WPP, especially when SBRs are scarce. However, our experiments suggest that training with bug reports from other available datasets, combined with BERT as the SBR predictor, could provide a viable solution.

\begin{tcolorbox}[yellowboxstyle]
    Relying solely on a single project for building SBR predictors for another is not advisable. However, our experiments indicate that using BRs from various projects with BERT as the predictor can be a viable solution.
\end{tcolorbox}

\section{Threats to validity}
\label{sec:ThreatsToValidity}
There are several threats to the validity of this study that we explain in the following.

\textbf{External validity.} Our analysis is limited to the five datasets Chromium, Derby, Ambari, Camel, and Wicket, which we did not update to the latest version to ensure a fair comparison with state-of-the-art studies. This may restrict the applicability of our results to current trends. We also used the FARSEC approach as a baseline but modified the existing implementation for functionality, potentially introducing discrepancies affecting result comparability. To generalize our findings, replication across multiple projects from different domains with labeled bug reports is required. Furthermore, while we focused on BERT as an encoder-only pre-trained model, other BERT variants could potentially offer improved performance.

\textbf{Internal validity.} There is a need to conduct multiple runs with varying data distributions for training and validation to ensure that performance metrics are reliable and not influenced by favorable or unfavorable data splits.
Further experimentation may lead to improved results. While our study may not achieve optimal SBR prediction, it highlights the benefits of data augmentation in addressing data imbalance. However, the question of how many datasets should be combined to build an effective model remains open for investigation.
We also did not perform an extensive hyperparameter search for model fine-tuning, which may lead to an underestimation of their performance. As a result, differences between the models could vary with different parameters, such as learning rates. However, since all models share a similar architecture, they are likely to perform best with comparable hyperparameters for the same task.

\textbf{Conclusion validity.}  Threats can impact the accuracy of our conclusions. Comparing learning system performance is challenging due to the variety of available metrics. In this study, we focus on the G-Measure for comparing automated classifications, as it combines recall and FPR into a single value.

\section{Conclusion}
\label{sec:Conclusion}

We conducted an extensive experimental evaluation of BERT and Random Forest (RF) for predicting Security Bug Reports (SBRs), focusing on both ``within-project'' and ``cross-project'' scenarios. Our experiments on five publicly available datasets show that RF outperforms BERT with a higher average G-measure for within-project predictions. Incorporating additional SBRs from various projects into the training data improves the average performance of both models. However, including a mix of security and non-security bug reports significantly reduces RF's average performance, yet boosts BERT to its highest average performance, which is much higher than RF's. In cross-project SBR prediction, BERT also achieves a remarkable G-measure, substantially higher than RF.

In future work, we plan to explore decoder-only transformer models, such as GPT, for SBR prediction.

\bibliographystyle{acm}
\bibliography{references}

\begin{thebibliography}{10}

\bibitem{CASMS}
Casms: Combining clustering with attention semantic model for identifying security bug reports.
\newblock {\em Information and Software Technology 147\/} (2022).

\bibitem{Zaher2021}
{\sc Alharthi, Z. S.~M., and Rastogi, R.}
\newblock An efficient classification of secure and non-secure bug report material using machine learning method for cyber security.
\newblock {\em Materials Today: Proceedings 37\/} (2021), 2507--2512.
\newblock International Conference on Newer Trends and Innovation in Mechanical Engineering: Materials Science.

\bibitem{Bruni2025}
{\sc Bruni, M., Gabrielli, F., Ghafari, M., and Kropp, M.}
\newblock Benchmarking prompt engineering techniques for secure code generation with gpt models.
\newblock In {\em Proceedings of the 2025 IEEE/ACM Second International Conference on AI Foundation Models and Software Engineering\/} (2025).

\bibitem{Noah2022}
{\sc B\"{u}hlmann, N., and Ghafari, M.}
\newblock How do developers deal with security issue reports on github?
\newblock In {\em Proceedings of the 37th ACM/SIGAPP Symposium on Applied Computing\/} (2022), SAC '22, p.~1580–1589.

\bibitem{10.1145/3528588.3528659}
{\sc Colavito, G., Lanubile, F., and Novielli, N.}
\newblock Issue report classification using pre-trained language models.
\newblock NLBSE '22, Association for Computing Machinery, p.~29–32.

\bibitem{10.1145/3643991.3644903}
{\sc Colavito, G., Lanubile, F., Novielli, N., and Quaranta, L.}
\newblock Leveraging gpt-like llms to automate issue labeling.
\newblock MSR '24, Association for Computing Machinery, p.~469–480.

\bibitem{BERTTransformers}
{\sc Devlin, J., Chang, M.-W., Lee, K., and Toutanova, K.}
\newblock Bert: Pre-training of deep bidirectional transformers for language understanding.
\newblock In {\em North American Chapter of the Association for Computational Linguistics\/} (2019).

\bibitem{devlinetal2019bert}
{\sc Devlin, J., Chang, M.-W., Lee, K., and Toutanova, K.}
\newblock {BERT}: Pre-training of deep bidirectional transformers for language understanding.
\newblock In {\em Proceedings of the 2019 Conference of the North {A}merican Chapter of the Association for Computational Linguistics: Human Language Technologies, Volume 1 (Long and Short Papers)\/} (Minneapolis, Minnesota, June 2019), J.~Burstein, C.~Doran, and T.~Solorio, Eds., Association for Computational Linguistics, pp.~4171--4186.

\bibitem{5463340}
{\sc Gegick, M., Rotella, P., and Xie, T.}
\newblock Identifying security bug reports via text mining: An industrial case study.
\newblock In {\em 2010 7th IEEE Working Conference on Mining Software Repositories (MSR 2010)\/} (2010), pp.~11--20.

\bibitem{Ghafari2017}
{\sc Ghafari, M., Gadient, P., and Nierstrasz, O.}
\newblock Security smells in android.
\newblock In {\em 2017 IEEE 17th International Working Conference on Source Code Analysis and Manipulation (SCAM)\/} (2017), pp.~121--130.

\bibitem{izadi2021predictingobjectivepriorityissue}
{\sc Izadi, M., Akbari, K., and Heydarnoori, A.}
\newblock Predicting the objective and priority of issue reports in software repositories, 2021.

\bibitem{LTRWES}
{\sc Jiang, Y., Lu, P., Su, X., and Wang, T.}
\newblock Ltrwes: A new framework for security bug report detection.
\newblock {\em Information and Software Technology 124\/} (2020), 106314.

\bibitem{reducefalsepositives}
{\sc Kharkar, A., Moghaddam, R.~Z., Jin, M., Liu, X., Shi, X., Clement, C., and Sundaresan, N.}
\newblock Learning to reduce false positives in analytic bug detectors.
\newblock In {\em Proceedings of the 44th International Conference on Software Engineering\/} (New York, NY, USA, 2022), ICSE '22, Association for Computing Machinery, p.~1307–1316.

\bibitem{DeeplearningBasedSoftwareBugClassification2024}
{\sc Meher, J.~P., Biswas, S., and Mall, R.}
\newblock Deep learning-based software bug classification.
\newblock {\em Information and Software Technology 166\/} (2024), 107350.

\bibitem{Ohiraetal}
{\sc Ohira, M., Kashiwa, Y., Yamatani, Y., Yoshiyuki, H., Maeda, Y., Limsettho, N., Fujino, K., Hata, H., Ihara, A., and Matsumoto, K.}
\newblock A dataset of high impact bugs: Manually-classified issue reports.
\newblock In {\em 2015 IEEE/ACM 12th Working Conference on Mining Software Repositories\/} (2015), pp.~518--521.

\bibitem{Pan_2024}
{\sc Pan, S., Luo, L., Wang, Y., Chen, C., Wang, J., and Wu, X.}
\newblock Unifying large language models and knowledge graphs: A roadmap.
\newblock {\em IEEE Transactions on Knowledge and Data Engineering 36}, 7 (July 2024), 3580–3599.

\bibitem{10.1145/3540250.3549156}
{\sc Pan, S., Zhou, J., Cogo, F.~R., Xia, X., Bao, L., Hu, X., Li, S., and Hassan, A.~E.}
\newblock Automated unearthing of dangerous issue reports.
\newblock In {\em Proceedings of the 30th ACM Joint European Software Engineering Conference and Symposium on the Foundations of Software Engineering\/} (New York, NY, USA, 2022), ESEC/FSE 2022, Association for Computing Machinery.

\bibitem{FARSEC}
{\sc Peters, F., Tun, T.~T., Yu, Y., and Nuseibeh, B.}
\newblock Text filtering and ranking for security bug report prediction.
\newblock {\em IEEE Transactions on Software Engineering 45}, 6 (2019), 615--631.

\bibitem{6860243}
{\sc Scandariato, R., Walden, J., Hovsepyan, A., and Joosen, W.}
\newblock Predicting vulnerable software components via text mining.
\newblock {\em IEEE Transactions on Software Engineering 40}, 10 (2014), 993--1006.

\bibitem{shu2019better}
{\sc Shu, R., Xia, T., Williams, L., and Menzies, T.}
\newblock Better security bug report classification via hyperparameter optimization, 2019.

\bibitem{Quentin22}
{\sc Sti\'{e}venart, Q., De~Roover, C., and Ghafari, M.}
\newblock Security risks of porting c programs to webassembly.
\newblock In {\em Proceedings of the 37th ACM/SIGAPP Symposium on Applied Computing\/} (New York, NY, USA, 2022), SAC '22, Association for Computing Machinery, p.~1713–1722.

\bibitem{vaswani2023attentionneed}
{\sc Vaswani, A., Shazeer, N., Parmar, N., Uszkoreit, J., Jones, L., Gomez, A.~N., Kaiser, L., and Polosukhin, I.}
\newblock Attention is all you need, 2023.

\bibitem{10.1109/TSE.2022.3178469}
{\sc von~der Mosel, J., Trautsch, A., and Herbold, S.}
\newblock On the validity of pre-trained transformers for natural language processing in the software engineering domain.
\newblock {\em IEEE Trans. Softw. Eng. 49}, 4 (Apr. 2023), 1487–1507.

\bibitem{Wetzels2023}
{\sc Wetzels, J., Dos~Santos, D., and Ghafari, M.}
\newblock Insecure by design in the backbone of critical infrastructure.
\newblock In {\em Proceedings of Cyber-Physical Systems and Internet of Things Week 2023\/} (New York, NY, USA, 2023), CPS-IoT Week '23, Association for Computing Machinery, p.~7–12.

\bibitem{6473768}
{\sc Wijayasekara, D., Manic, M., Wright, J.~L., and McQueen, M.}
\newblock Mining bug databases for unidentified software vulnerabilities.
\newblock In {\em 2012 5th International Conference on Human System Interactions\/} (2012), pp.~89--96.

\bibitem{wolfetal2020transformers}
{\sc Wolf, T., Debut, L., Sanh, V., Chaumond, J., Delangue, C., Moi, A., Cistac, P., Rault, T., Louf, R., Funtowicz, M., Davison, J., Shleifer, S., von Platen, P., Ma, C., Jernite, Y., Plu, J., Xu, C., Le~Scao, T., Gugger, S., Drame, M., Lhoest, Q., and Rush, A.}
\newblock Transformers: State-of-the-art natural language processing.
\newblock In {\em Proceedings of the 2020 Conference on Empirical Methods in Natural Language Processing: System Demonstrations\/} (Online, Oct. 2020), Q.~Liu and D.~Schlangen, Eds., Association for Computational Linguistics, pp.~38--45.

\bibitem{9371393}
{\sc Wu, X., Zheng, W., Xia, X., and Lo, D.}
\newblock Data quality matters: A case study on data label correctness for security bug report prediction.
\newblock {\em IEEE Transactions on Software Engineering 48}, 7 (2022).

\bibitem{10.1007/s10515-014-0162-2}
{\sc Xia, X., Lo, D., Shihab, E., Wang, X., and Zhou, B.}
\newblock Automatic, high accuracy prediction of reopened bugs.
\newblock {\em Automated Software Engg. 22}, 1 (Mar. 2015), 75–109.

\end{thebibliography}
\end{document}